\documentclass[12pt]{article}
\usepackage{a4}
\usepackage{amsfonts}
\usepackage{amsmath,amssymb}

\usepackage{graphicx}

\makeatother

\def\un#1{\relax\ifmmode\@@underline#1\else
        $\@@underline{\hbox{#1}}$\relax\fi}

\let\du=\du                     

\def\a{\alpha}
\def\b{\beta}
\def\c{\chi}
\def\d{\delta}

\def\f{\phi}
\def\g{\gamma}
\def\h{\eta}

\def\j{\psi}

\def\l{\lambda}
\def\m{\mu}
\def\n{\nu}

\def\p{\pi}
\def\q{\theta}
\def\r{\rho}
\def\s{\sigma}

\def\x{\xi}

\def\D{\Delta}

\def\G{\Gamma}

\def\ve{\varepsilon}

\def\bo{{\raise-.3ex\hbox{\large$\Box$}}}               
\def\pa{\partial}                                       
\def\TH{{\raise.2ex\hbox{$\displaystyle \bigodot$}\mskip-4.7mu \llap H \;}}
\def\face{{\raise.2ex\hbox{$\displaystyle \bigodot$}\mskip-2.2mu \llap {$\ddot
        \smile$}}}                                      
\def\dg{\sp\dagger}                                     

\def\sp#1{{}^{#1}}                              
\def\VEV#1{\left\langle #1\right\rangle}        
\def\leftrightarrowfill{$\mathsurround=0pt \mathord\leftarrow \mkern-6mu
        \cleaders\hbox{$\mkern-2mu \mathord- \mkern-2mu$}\hfill
        \mkern-6mu \mathord\rightarrow$}
\def\dvec#1{\vbox{\ialign{##\crcr
        \leftrightarrowfill\crcr\noalign{\kern-1pt\nointerlineskip}
        $\hfil\displaystyle{#1}\hfil$\crcr}}}           

\def\frac#1#2{{\textstyle{#1\over\vphantom2\smash{\raise.20ex
        \hbox{$\scriptstyle{#2}$}}}}}                   
\def\sfrac#1#2{{\vphantom1\smash{\lower.5ex\hbox{\small$#1$}}\over
        \vphantom1\smash{\raise.4ex\hbox{\small$#2$}}}} 
\def\bfrac#1#2{{\vphantom1\smash{\lower.5ex\hbox{$#1$}}\over
        \vphantom1\smash{\raise.3ex\hbox{$#2$}}}}       
\def\afrac#1#2{{\vphantom1\smash{\lower.5ex\hbox{$#1$}}\over#2}}    
\def\on#1#2{\mathop{\null#2}\limits^{#1}}               
\def\bvec#1{\on\leftarrow{#1}}                  

\def\[{\lfloor{\hskip 0.35pt}\!\!\!\lceil}
\def\]{\rfloor{\hskip 0.35pt}\!\!\!\rceil}

\def\du#1#2{_{#1}{}^{#2}}
\def\ud#1#2{^{#1}{}_{#2}}

\def\udu#1#2#3{^{#1}{}_{#2}{}^{#3}}

\def\ha{{\fracmm12}}
\def\tr{{\rm tr}}

\def\un{\underline}
\def\fracmm#1#2{{{#1}\over{#2}}}

\def\low#1{{\raise -3pt\hbox{${\hskip 0.75pt}\!_{#1}$}}}

\newskip\humongous \humongous=0pt plus 1000pt minus 1000pt
\def\caja{\mathsurround=0pt}
\def\eqalign#1{\,\vcenter{\openup2\jot \caja
        \ialign{\strut \hfil$\displaystyle{##}$&$
        \displaystyle{{}##}$\hfil\crcr#1\crcr}}\,}
\newif\ifdtup

\newcommand{\be}{\begin{equation}}
\newcommand{\ee}{\end{equation}}
\newcommand{\nbe}{\begin{equation*}}
\newcommand{\nee}{\end{equation*}}

\newcommand{\lb}{\label}

\begin{document}

\thispagestyle{empty}

{\hbox to\hsize{
\vbox{\noindent January 2011 \hfill revised version  }}}

\noindent
\vskip2.0cm
\begin{center}

{\large\bf Higher-derivative gauge interactions of 
\vglue.1in Bagger-Lambert-Gustavsson theory in $N=1$ superspace~\footnote{
Supported in part by the Japanese Society for Promotion of Science (JSPS) and 
the German \newline ${~~~~~}$ Max-Planck-Society (MPI f\"ur Physik, 
M\"unchen)}}
\vglue.3in

Sergei V. Ketov~${}^{a,b}$ and Shutaro Kobayashi${}^a$ 
\vglue.1in

${}^a$ {\it Department of Physics, Tokyo Metropolitan University, Japan}\\
${}^b$ {\it IPMU, University of Tokyo, Japan}
\vglue.1in
ketov@tmu.ac.jp, kobayashi-shutaro@ed.tmu.ac.jp
\end{center}

\vglue.3in

\begin{center}
{\Large\bf Abstract}
\end{center}
\vglue.1in

\noindent We study the structure of the gauge sector of the Bagger-Lambert-Gustavsson (BLG) 
theory in the form proposed by van Raamsdonk, adapted to 3D, $N=1$ superspace. By using the
novel Higgs mechanism proposed by Mukhi and Papageorgakis, we derive the manifestly $N=1$ 
supersymmetric higher-order terms (beyond the supersymmetric Yang-Mills action) that follow from 
the BLG theory in its expansion with respect to the inverse gauge coupling constant squared.
We find that all those terms have at least one anti-commutator of the super-YM field strength 
superfields as a factor, and thus are reducible to the SYM terms with the higher (spacetime) 
derivatives. 

\noindent 

\newpage

\section{Introduction}

Unlike the notion of the {\it Abelian} Born-Infeld (BI) action which is well defined,
the notion of a {\it Non-Abelian} Born-Infeld (NBI) action is rather misleading because
it is dependent upon a perturbation theory, and it has to include the terms depending 
upon the derivatives of the fields. So it may be better to talk about particular deformation
of the Yang-Mills theory by specifying the underlying theory that leads to the higher-order terms.  
The same observations equally apply to the supersymmetric extensions of the NBI-type actions 
(see eg., refs.~\cite{ark,mf} for reviews). For instance, a (perturbatively defined) NBI 
action naturally arises as the effective action from the open superstring scattering 
amplitudes \cite{kita,brus,ital}, whereas the Abelian BI action arises as the low-energy
effective action of a single D-brane \cite{lei}. The effective action of  multiple 
(coinciding) D-branes also contains an NBI action \cite{witten}, though many attempts to 
explicitly construct such NBI action were not very successful (see eg., ref.~\cite{belg,hlw}). 
Another NBI action is supposed to arise as part of the effective action of multiple 
M-branes.~\footnote{As regards some specific NBI proposals, see also 
refs.~\cite{tsey,ket2,nbi,sis}.}

Yet another ambiguity in defining an NBI action is related to the fact that (unlike the abelian
case) allowing large values of the non-abelian YM field necessarily implies allowing 
large values of its derivatives because of the identity
\be  \lb{ymi}
  \[ F_{\m\n}, F_{\l\r} \] = D_{[\m}D_{\n]} F_{\l\r}  
\ee
The same conclusion arises by requiring the absence of a formation of black holes in an
NBI theory \cite{bilal}. In other words, the presense of the higher-derivative terms in
any NBI action is inevitable and model-dependent.

More recently, multiple M2-branes were investigated in the context of BLG theory 
\cite{bl,gu} (see also the ABJM theory \cite{abjm} and ref.~\cite{cop} for a recent review).
Though the original BLG theory may be only applicable to two coinciding M2-branes, a variation 
of the Higgs mechanism arising in the BLG and ABJM theories \cite{mukhi} (see also 
ref.~\cite{nic}) directly gives rise to a perturbative expansion in terms of the inverse YM 
coupling constant squared, $g^{-2}_{\rm YM}$ (or, equivalently, in terms of the inverse vacuum 
expectation value of the Higgs field, $\VEV{X}^{-1}$), before the $\a'$-corrections are taken 
into account. As a result, a new higher-order action arises that is truly non-abelian. That 
higher-order expansion is complementary to the standard $\a'$-expansion, while they are truly
independent. The $\a'$-corrections are inherent to any D-brane action, whereas the 
$g^{-2}_{\rm YM}$ corrections are inherent to a particular background that $D2$-branes are probing.

The original BLG and ABJM actions have a high amount of supersymmetry, but we are going to 
concentrate on a particular (gauge) sector  of those supersymmetric gauge theories  by using 
simple (or $N=1$) superfields in three dimensions (3D). The superfield description of the BLG 
and ABJM theories  in 3D, $N=1$ superspace was given in 
ref.~\cite{ana}.~\footnote{See refs.~\cite{rus,swe,wim} for 
the BLG/ABJM theories in terms of extended superfields in 3D.}

Our paper is organized as follows. Sec.~2 is our setup devoted to a superspace description
of 3D supersymmetric gauge theories. In sect.~3 we introduce our model as part of the
BLG/ABJM theory. A calculation of the higher-order (and higher-derivative) gauge terms  
resulting from the Higgs mechanism is given is Sec.~4. Our conclusion is Sec.~5. Our
notation and superspace conventions are collected in Appendices A and B.

\section{Setup}

An $N=1$ supersymmetric non-Abelian gauge theory in three spacetime dimensions is 
defined in flat 3D superspace $z^A=(x^{\m},\q_{\a})$ via the Lie algebra-valued 
gauge- and super-covariant derivatives~\footnote{Our notation and conventions are
collected in Appendices A and B.}
\be \lb{covder}
\nabla_A = D_A + i \G_A 
\ee
subject to the off-shell superfield constraints \cite{sspace}
\be \lb{scon}
 \{ \nabla_{\a}, \nabla_{\b} \} = -2i\nabla_{a\b}
\ee 
Equivalently, eq.~(\ref{scon}) means that the vector gauge connection is not 
independent but can be writen down in terms of the spinor gauge connections as
follows:~\footnote{All (anti)symmetrizations of indices are defined without a weight 
factor.} 
\be \lb{scon2}
\G_{\a\b} = \frac{i}{2} \left( D_{(\a}\G_{\b)}+i\{ \G_{\a},\G_{\b} \} \right)
\ee
The supergauge connection $\G_A$ belongs to the adjoint representation of the gauge
group. Under the supergauge transformations with the gauge Lie algebra-valued 
 parameter $K$ the connection transforms as
\be \lb{cotr}
\d \G_A = \nabla_A K = D_A K + i \[ \G_A, K \] 
\ee

The superspace Bianchi identities 
\be \lb{sbia}
 \[ \nabla_{\a}, \{ \nabla_{\b},\nabla_{\g} \} \] +
 \[ \nabla_{\b}, \{ \nabla_{\g},\nabla_{\a} \} \] +
 \[ \nabla_{\g}, \{ \nabla_{\a},\nabla_{\b} \} \] =0
\ee
subject to the conventional superspace constraints (\ref{scon}) imply \cite{sspace}
\be \lb{eqw} 
 \[ \nabla_{\a}, \nabla_{\b\g} \] = -\ve_{\a(\b}W_{\g)} 
\ee
where we have introduced the non-Abelian supercovariant superfield strtength
\be \lb{sstr}
W_{\a} = \frac{1}{2} D^{\b}D_{\a}\G_{\b} +\frac{i}{2}\[ \G^{\b},D_{\b}\G_{\a} \]
-\frac{1}{6}\[ \G^{\b},\{ \G_{\b},\G_{\a} \} \] 
\ee
obeying the geometrical (off-shell) constraint
\be \lb{bia2}
\nabla^{\a}W_{\a}=0
\ee
 
The spinor superfield $\G_{\a}$ can be expanded in terms of its field
components as
\be \lb{fexp}
\G_{\a} = \c_{\a} + \frac{1}{2}\q_{\a}B + (\g^{\m}\q)_{\a}A_{\m} +
i\q^2 \left[ \l_{\a} -\frac{1}{2}(\g^{\m} \pa_{\m}\c)_{\a} \right]
\ee 
The supergauge transformations (\ref{cotr}) can be used to impose a Wess-Zumino
(WZ) gauge 
\be \lb{wzg}
\c_{\a} =B =0
\ee
The remaining non-Abelian fields $A_{\m}$ and $\l_{\a}$ can be identified as a
Yang-Mills (YM) gauge field and a gaugino, resectively. The covariant superfield 
strength $W_{\a}$ takes the form
\be \lb{strcom}
W_{\a} = i\l_{\a} + \frac{i}{2}(\g^{\m}\g^{\n}\q)_{\a} F_{\m\n} 
-\frac{1}{2}\q^2(\g^{\m}D_{\m}\l)_{\a}
\ee
in terms of the conventional YM field strength $F_{\m\n}=\pa_{\m}A_{\n}
-\pa_{\n}A_{\m} +i\[ A_{\m},A_{\n} \]$ and the (Dirac) covariant derivative
$(D_{\m}\l)_{\a} = \pa_{\m}\l_{\a} + i \[ A_{\m},\l_{\a} \] $.

The exist {\it two} natural (supersymmetric and gauge-invariant) actions in 3D
superspace: the super-YM action \cite{sspace}
\be \lb{sym}
S_{\rm sYM}=\fracmm{1}{8g_{\rm YM}^2} \int d^3x d^2\q\, \tr\left(
W^{\a}W_{\a}\right) = \fracmm{1}{g^2_{\rm YM}}\int d^3x\, \tr\left(
-\frac{1}{4}F_{\m\n}F^{\m\n} - \frac{i}{2}\l\g^{\m}D_{\m}\l \right)  
\ee
and the super-Chern-Simons (super-CS) action \cite{sspace,nishig}
\be
\eqalign{ 
S_{\rm sCS}  &~ = \fracmm{1}{8f_{\rm CS}} \int d^3x d^2\q\, \tr \left(
i\G^{\a}W_{\a}+\frac{1}{6}\{\G^{a},\G^{\b}\}D_{\b}\G_{\a} +\fracmm{i}{12}
\{ \G^{\a},\G^{\b}\}\{\G_{\a},\G_{\b}\}\right) \cr 
 &~ = \fracmm{1}{2f_{\rm CS}}\int d^3x\, \tr \left(
-i(\l\l) + \ve^{\m\n\r}\left[ A_{\m}\pa_{\n}A_{\r} +\frac{2i}{3}A_{\m}A_{\n}A_{\r}
\right] \right)  \lb{scs} \cr}
\ee
It is worth mentioning that the last term of the super-CS action in 3D superspace
vanishes in the WZ-gauge. As is well-known, the CS coupling constant $f_{\rm CS}$
gets quantized as $f_{\rm CS}=2\p/k$ where $k\in {\bf Z}$ \cite{witt}.

\section{Our model}

The 3D model we consider is given by an $N=1$ supersymmetric gauge field theory with
the gauge group $G\times G$ and the superfield action
\be \lb{action}
 S = S_{\rm matter} + S_{\rm CS}^{(1)} - S_{\rm CS}^{(2)}
\ee
The matter action is given by
\be \lb{mata}
 S _{\rm matter} = \frac{1}{4} \int d^3x d^2\q\, \tr\left(\nabla^{\a}X^{\dg}
\nabla_{\a}X\right)
\ee
with the matter superfield $X$ in the bi-fundamental representation of the gauge group,
\be \lb{bif}
\nabla_{\a} X = D_{\a}X +i\G^{(1)}_{\a}X -i X \G^{(2)}_{\a} 
\ee
where we have introduced the gauge connections $\G^{(1)}_{\a}$ and $\G^{(2)}_{\a}$ for
each factor $G$ in the gauge group $G\times G$. The $S^{(1)}_{\rm CS}$  and 
$S^{(2)}_{\rm CS}$ are the super-CS actions (\ref{scs}) for each gauge factor $G$.
  
In terms of the field components the action (\ref{action}) represents the gauge part of the
ABJM or BLG actions when $G=U(N)$ and $G=SU(2)$, respectively, in the form proposed in
ref.~\cite{rams}. Since our purpose is to generate a supersymmetric higher-order action for 
the super-YM fields, we consider only the relevant terms in what follows. 

By giving an expectation value to the scalar $X$ as
\be \lb{expect}
\VEV{X} ={\rm const.} \neq 0
\ee
it is possible to spontaneously break the gauge group $G\times G$ to its diagonal subgroup
$G_{\rm diag}=G$, as in ref.~\cite{mukhi}. Then the gauge fields 
\be \lb{unb} 
\G_{\a}=\frac{1}{2}\left( \G^{(1)}_{\a}+\G^{(2)}_{\a}\right)
\ee
are going to be associated with the {\it unbroken} gauge symmetry $G_{\rm diag}$, 
whereas the  rest of the gauge fields 
\be \lb{brok} 
\D_{\a}=\frac{1}{2}\left( \G^{(1)}_{\a}-\G^{(2)}_{\a}\right)
\ee
are going to be associated with the {\it broken} gauge symmetry.

The $\D$-dependent part of the Lagrangian in eq.~(\ref{action}) is given by
\be \lb{ddep}
L(\D,W) =\VEV{X}^2 \tr(\D^{\a}\D_{\a}) +\fracmm{i}{2f_{\rm CS}}
\tr(\D^{\a}W_{\a}) -\fracmm{1}{12f_{\rm CS}} \tr(\{\D^{\a},\D^{\b}\}\nabla_{\b}\D_{\a})
\ee
where $W_{\a}$ is the YM superfield strength (\ref{strcom}) and
\be \lb{newco}
\nabla_{\b}\D_{\a} = D_{\b}\D_{\a}+i \{ \G_{\b},\D_{\a} \}
\ee
 As is clear from  eq.~(\ref{ddep}), the supefields $\D_{\a}$ are not propagating but 
represent the auxiliary degrees of freedom that can be eliminated via their
non-dynamical equations of motion,
\be \lb{auxe}
 \D_{\a} = -\fracmm{i}{4\VEV{X}^2f_{\rm CS}}W_{\a} - \fracmm{1}{8\VEV{X}^2f_{\rm CS}}
\[\D_{\b},\nabla^{\b}\D_{\a} \] 
\ee

By using the equations collected in Sec.~2 and the Appendices A and B, we find the 
bosonic (gauge) part of eq.~(\ref{ddep}) in the form
\be \lb{dbos}
L_{\rm bos}(B,F) = 4\VEV{X}^2 \tr (B^{\m}B_{\m}) +
\fracmm{1}{f_{\rm CS}}\ve^{\m\n\r}\tr \left( B_{\m}F_{\n\r}
+\frac{2i}{3}B_{\m}B_{\n}B_{\r}\right)
\ee
in terms of the vector gauge field component $B_{\m}$ of $\D_{\a}$, and the YM field
strength $F_{\m\n}$. Equation (\ref{dbos}) agrees with the known results of
 refs.~\cite{mukhi,china}.

\section{Higher-derivative super-Yang-Mills terms}

Equation (\ref{ddep}) does lead to the truly Non-Abelian deformation of the YM action by 
the higher-order terms with the higher derivatives. Though it is impossible to solve 
eq.~(\ref{auxe}) for $\D_{\a}$ in a finite explicit form, it is always possible (and easy) to 
get an iterative solution up to any given order in $W$. Substituting the iterative solution 
back into the action  (\ref{ddep}) and using the identity (\ref{bia2}), we find
\be
\eqalign{
L(W)  &~ = \fracmm{1}{2^4\VEV{X}^2f^2_{\rm CS}}\tr\left(W^{\a}W_{\a}\right) -
\fracmm{i}{3\cdot 2^8\VEV{X}^6f^4_{\rm CS}}\tr\left( \{W^{\a}, W^{\b}\} 
\nabla_{\b}W_{\a}\right)\cr
&~ - \fracmm{1}{2^{14} \VEV{X}^{10}f^6_{\rm CS}}\tr\left( \{W^{\a}, W^{\b} \} 
\nabla_{\b}\nabla^{\g} \{ W_{\g},W_{\a} \} \right) +
{\cal O}(\VEV{X}^{-14}f^{-8}_{\rm CS}) \lb{nbie} \cr}
\ee
The first term just represents the super-Yang-Mills Lagrangian in superspace (Sec.~2), 
whereas the other terms have the spinorial covariant derivatives of the YM superfield 
strength $W$. The peculiar feature of those extra terms is the presence of the 
anti-commutator 
\be \lb{antico}  
\{ W_{\a}, W_{\b} \} = -\frac{1}{6} \[ \nabla^{\g},\nabla_{\g(\a}\] W_{\b)}
\ee
that has the spacetime derivative of $W$. It means that all the extra terms beyond the
super-Yang-Mills term in eq.~(\ref{nbie}) give rise to the spacetime higher-derivative 
contributions with respect to the YM field strength in components. The same conclusion 
also follows from the compoment form of the anticommutator (\ref{antico}),
\be \lb{compcom}
\{ W_{\a}, W_{\b} \} = -\frac{1}{4} (\g^{\m}\g^{\n}\q)_{\a} (\g^{\r}\g^{\s}\q)_{\b}
\[ F_{\m\n},F_{\r\s} \] +{\rm ~fermionic~~terms.}
\ee

To the end of this section we explicitly derive the bosonic (YM) contributions out of the 
higher-order terms in eq.~(\ref{nbie}). The good starting point is eq.~(\ref{dbos}).
Varying it with respect to $B^{\m}$ yields
\be \lb{iter}
B^{\m} = - \fracmm{1}{g}\ve^{\m\n\r}F_{\n\r} - \fracmm{2i}{g}\ve^{\m\n\r}B_{\n}B_{\r}  
\ee
where $\frac{1}{2}\ve^{\m\n\r}F_{\n\r}={}^*F^{\m}$ and $g=8\VEV{X}^2f_{\rm CS}$. 
Equation (\ref{iter}) is well suitable for doing iterations with
respect to $B^{\m}$ or expanding its solution in terms of the inverse powers of $g$,
\be \lb{expan}
B^{\m} = -\sum^{+\infty}_{n=0} \fracmm{1}{g^n}S^{\m}_n
\ee
where 
\be \lb{itsb}
S^{\m}_0=\left(\fracmm{2}{g}\right){}^*F^{\m}\qquad {\rm and}\qquad 
S^{\m}_{k+1}=i\ve^{\m\n\r}\sum _{n+m=k \atop n,m\geq 0}\[ S_{n\n},S_{m\r} \]
\ee
Substituting the solution (\ref{itsb}) back into the action (\ref{dbos}) yields
the NBI action 
\be \lb{expl}
 L(F) =   \sum^{+\infty}_{n=2} L^{(n)} 
\ee
where
\be \lb{l2}
L^{(2)} = -\fracmm{1}{2^3\VEV{X}^2f^2_{\rm CS}}\tr \left( F^{\m\n}F_{\m\n}\right)~, 
\ee
\be \lb{l3}
L^{(3)} = -\fracmm{i}{3\cdot 2^6\VEV{X}^6f^4_{\rm CS}}\tr \left( \[ F^{\m\n},F_{\n\r} \]
F\ud{\r}{\m}\right)~,
\ee
\be \lb{l4}
L^{(4)} = \fracmm{1}{2^{13}\VEV{X}^{10}f^6_{\rm CS}}\tr \left( \[F_{\m\n},F_{\r\s} \]
\[F^{\m\n},F^{\r\s} \]\right)~,
\ee
\be \lb{l5}
L^{(5)} = \fracmm{i}{2^{15}\VEV{X}^{14}f^8_{\rm CS}}\tr \left( \[F_{\m\n},F_{\r\s} \]
\[F^{\r\l},F\du{\l}{\s} \] F^{\m\n}\right)~,
\ee
\be \lb{6}
\eqalign{
L^{(6)} = ~&~ -\fracmm{1}{2^{21}\VEV{X}^{18}f^{10}_{\rm CS}}\tr \left( 
\[ \[F_{\m\n},F_{\r\s} \],F^{\r\s} \] \[ \[F^{\m\n},F^{\l\eta} \], F_{\l\eta}\]\right) \cr
~&~ - \fracmm{1}{3\cdot 2^{20}\VEV{X}^{18}f^{10}_{\rm CS}}\tr 
\left( \[F^{\m\n},F^{\r\s} \]\[ F_{\r\s},F_{\l\eta} \] \[F^{\l\eta},F_{\m\n} \] \right)~.
\cr}
\ee

It is instructive to specify those equations to the case of the $G=SU(2)$ gauge group with
$B_{\m}=B_{\m}^a\s^a$ and $F_{\m\n}=F_{\m\n}^a\s^a$, where $\s^a$ are Pauli matrices, 
$a=1,2,3$. Equation (\ref{dbos}) then takes the form 
\be \lb{dbos2}
L(B,F) = 8\VEV{X}^2B^{\m}\cdot B_{\m} +\fracmm{2}{f_{\rm CS}}\ve^{\m\n\r}B_{\m}
\cdot F_{\n\r} - \fracmm{4}{3f_{\rm CS}}\ve^{\m\n\r}
\left(B_{\m}\times B_{\n}\right)\cdot B_{\r}
\ee
where we have introduced the usual scalar and vector products of $SU(2)$ vectors 
in three dimensions,
\be \lb{2pro}
A\cdot B = A^aB^a\qquad {\rm and} \qquad (A\times B)^a = \ve^{abc}A^bB^c 
\ee
After substituting the iterative solution of the $B$-equation of motion back into the 
Lagrangian (\ref{dbos2}), we find
\be \lb{2l2}
L^{(2)} = -\fracmm{1}{4\VEV{X}^2f^2_{\rm CS}} F^{\m\n}\cdot F_{\m\n}~~,
\ee
\be \lb{2l3}
L^{(3)} = \fracmm{1}{3\cdot 2^4\VEV{X}^6f^4_{\rm CS}} \left( F^{\m\n}\times 
F_{\n\r}\right)\cdot F\ud{\r}{\m}= \fracmm{1}{3\cdot 2^4\VEV{X}^6f^4_{\rm CS}} 
\ve^{abc}F^{\m\n a}F^b_{\n\r} F\udu{\r}{\m}{c}~~,
\ee
\be \lb{2l4}
\eqalign{
L^{(4)} ~&~ =
-\fracmm{1}{2^{10}\VEV{X}^{10}f^6_{\rm CS}} \left( F^{\m\n}\times F^{\r\s}\right)
\cdot \left( F_{\m\n}\times F_{\r\s}\right) \cr
~&~  =  \fracmm{1}{2^{10}\VEV{X}^{10}f^6_{\rm CS}} \left\{ 
\left( F^{\m\n}\cdot F^{\r\s}\right) \left( F_{\m\n}\cdot F_{\r\s}\right) -
 \left( F^{\m\n}\cdot F_{\m\n}\right) \left( F^{\r\s}\cdot F_{\r\s}\right)
\right\}~~, \cr}
\ee
\be \lb{2l5}
\eqalign{
L^{(5)} ~&~ = \fracmm{1}{2^{12}\VEV{X}^{14}f^8_{\rm CS}} \left[
\left(  F_{\m\n}\times F_{\r\s} \right) \times \left(
 F^{\r\l}\times F\du{\l}{\s} \right) \right] \cdot F^{\m\n} \cr
~&~ = \fracmm{1}{2^{12}\VEV{X}^{14}f^8_{\rm CS}}\ve^{abc}
\left( F^a_{\m\n}F^{\n\r b}F\du{\r}{\m c}F^d_{\s\l}F^{\s\l d}
+F^{\m\n a}F\udu{\s}{\r}{b}F^{\r\l c}F^d_{\m\n}F^d_{\s\l}\right)~~,
\cr}
\ee
\be \lb{2l6}
\eqalign{
L^{(6)} = ~&~  -\fracmm{1}{2^{16}\VEV{X}^{18}f^{10}_{\rm CS}} \left[
\left(  F_{\m\n}\times F_{\r\s} \right) \times F^{\r\s}\right]\cdot
 \left[ \left(  F^{\m\n}\times F^{\l\eta} \right) \times F_{\l\eta}
\right] \cr
~&~  -\fracmm{1}{3\cdot 2^{16}\VEV{X}^{18}f^{10}_{\rm CS}} \left[
\left(  F_{\m\n}\times F_{\r\s} \right) \times \left( F^{\r\s}\times
F^{\l\eta}\right)\right] \cdot \left(  F_{\l\eta}\times F^{\m\n} \right) 
\cr
~&~ = -\fracmm{1}{3\cdot 2^{16}\VEV{X}^{18}f^{10}_{\rm CS}} \left\{
4 \left(F^{\m\n}\cdot F_{\m\n}\right) \left(F^{\r\s}\cdot F_{\r\s}\right)
\left(F^{\l\eta}\cdot F_{\l\eta}\right)\right. \cr
~&~ ~~~~~~~~~~~~~~~~~~~~~~~~~~+5
\left(F^{\m\n}\cdot F^{\r\s}\right)\left(F_{\r\s}\cdot F_{\l\eta}\right)\left(F^{\l\eta}
\cdot F_{\m\n}\right)  \cr
~&~ \left. ~~~~~~~~~~~~~~~~~~~~~~~~~-9 
\left( F^{\m\n}\cdot F_{\m\n}\right)\left(F^{\r\s}\cdot F^{\l\eta}\right)
\left( F_{\r\s}\cdot F_{\l\eta}\right)\right\}~~.\cr}
\ee

\section{Conclusion}

It follows from eqs.~(\ref{l2}) and (\ref{2l2}) that the YM coupling constant
is given by
\be \lb{ymc}
g_{\rm YM } = \VEV{X} f_{\rm CS}
\ee

All the terms we found beyond the (s)YM action do not have the Abelian analogue -- 
they simply vanish in the Abelian case. Moreover, each  higher-order term has at least 
one (anti)commutator of the (s)YM field strengths. It also implies that all those terms 
beyond the sYM action can be rewritten to the form with the spacetime derivatives of the 
(s)YM field strength. Our results in superspace agree with the earlier observations in
terms of the field components \cite{china}. It implies that the action structure of a single 
(Abelian) M2-brane and that of multiple (Non-Abelian) M2-branes are very different. For example,
the higher-order gauge interactions considered in this paper for multiple D2-branes have no
counterpart for a single D2-brane.  

To avoid confusion, we would like to stress again that because of eq.~(\ref{ymc}) the non-abelian
expansion in powers of the inverse expectation value of the Higgs field in Sec.~4 is, in fact, an
expansion in terms of the inverse YM coupling constant squared (or, equivalently, in terms of the
inverse Chern-Simons parameter $f_{\rm CS}$), and it has nothing to do with the perturbative 
expansion of the DBI action of multiple D2-branes in terms of the string constant $\a'$. In the 
(M-theory) brane realization of the Mukhi-Papageorgakis mechanism the relevant geometry is given 
by a cone whose opening angle is proportional to the Chern-Simons parameter $f_{\rm CS}$, while the 
relevant geometry of the corresponding D2-branes is that of a thin cylinder \cite{abjm,cop}. Therefore, 
the corrections in terms of the inverse of $f_{\rm CS}$ come from moving the M2-branes away from the 
singularity at the top of the cone, and they describe deviations of the cylinder geometry from that of 
the cone.~\footnote{The authors appreciate this comment of the referee.} 

Our explicit results may also be considered as the supersymmetric generalization of the earlier results 
about the BLG theory \cite{china} obtained in the bosonic case. 

\section*{Acknowledgements}

The authors thank S.J. Gates Jr., D. Tsimpis, R. Wimmer and B. Zupnik for correspondence, and one of
the referees for his valuable comments. One of the authors (SVK) is grateful to the 
Max-Planck-Institute of Physics in Munich and  the Centre for Theoretical Physics in Marseille 
for kind hospitality  extended to him during preparation of this paper. 

\section*{Appendix A: spacetime notation}

Our 3D metric is $\h_{\m\n}={\rm diag}(+,-,-)$, where $\m,\n=0,1,2$. We use the
$SL(2,{\bf R})$ notation for Lorentz transformations in 3D. In particular, 3D 
spinors belong to the fundamental representation ${\bf 2}$ of $SL(2,{\bf R})$.We use 
the lower case {\it early} Greek letters for {\it spinor} indices, and the lower case 
{\it middle} Greek letters for {\it vector} indices. We also avoid explicit writing 
of spinor indices whenever it does not lead to confusion.

As far as the $SL(2,{\bf R})$ generators are concerned, we choose
\be \lb{gener}
T^0=\frac{1}{2} \left( \begin{array}{cc} 0 & 1 \\ -1 & 0 \end{array} \right)~,\quad
T^1=\frac{1}{2} \left( \begin{array}{cc} 0 & 1 \\ 1 & 0 \end{array} \right)~,\quad
T^2=\frac{1}{2} \left( \begin{array}{cc} 1 & 0 \\ 0 & -1 \end{array} \right)
\ee 
with the (anti)commutation relations 
\be\lb{commm}
\[ T^{\m},T^{\n} \] =f\ud{\m\n}{\r}T^{\r}~,\quad
 \{ T^{\m},T^{\n} \} =-\ha \h^{\m\n}
\ee
and the Killing form
\be \lb{kill}
\tr \left( T^{\m}T^{\n}\right) =-\ha \h^{\m\n} 
\ee

The 3D vector indices are raised and lowered with $\h^{\m\n}$ and $\h_{\m\n}$. It
follows
\be \lb{antis}
 f^{\m\n\r} =\h^{\r\s}f\ud{\m\n}{\s}=-\ve^{\m\n\r} 
\ee
where $\ve^{\m\n\r}$ is 3D Levi-Civita symbol, $\ve^{012}=1$. 

We  define 3D (Dirac) gamma matrices $(\g_{\m})\du{\a}{\b}$ by
\be \lb{gamm}
\g_{\m} = 2T_{\m}~,\qquad \{ \g_{\m},\g_{\n} \} =-2\h_{\m\n} 
\ee
so that
\be \lb{gammp}
(\g^{\m})\du{\a}{\b}(\g_{\m})\du{\g}{\d}=-2\d^{\d}_{\a}\d^{\b}_{\g}+
\d^{\b}_{\a}\d^{\d}_{\g}
\ee
In addition, we have
\be \lb{gammi1}
\tr(\g_{\m})=0~,\qquad \tr(\g_{\m}\g_{\n})=-2\h_{\m\n} \ee
and
\be \lb{gammi2}
\tr(\g^{\m}\g^{\n}\g^{\r})=2\ve^{\m\n\r}~,\quad 
\tr(\g^{\m}\g^{\n}\g^{\r}\g^{\s})=2\h^{\m\n}\h^{\r\s}-2\h^{\m\r}\h^{\s\n}
+2\h^{\m\s}\h^{\n\r}
\ee

The spinor indices are raised and lowered by the real antisymmetric symbols
$\ve^{\a\b}$ and $\ve_{\a\b}$ with
\be \lb{ves}
\ve^{\a\b} = \left( \begin{array}{cc} 0 & -1 \\ 1 & 0 \end{array} \right)~~,\qquad
\ve_{\a\b} = \left( \begin{array}{cc} 0 & 1 \\ -1 & 0 \end{array} \right)
\ee
so that
\be \lb{vei}
\ve^{\a\b}\ve_{\g\d} = \d^{\a}_{\d}\d^{\b}_{\g}-\d^{\a}_{\g}\d^{\b}_{\d}
\ee
The book-keeping notation
\be \lb{bkeep}
(\q\j)=\q^{\a}\j_{\a}\quad {\rm and} \quad (\j\j)=\j^2
\ee
is used for the scalar products of spinors. 

A conversion between a vector $V^{\m}$ and the associated bi-spinor $V\ud{\a}{\b}$ 
is given by
\be \lb{conv}
 V^{\m} =\frac{1}{2}(\g^{\m})\du{\a}{\b}V\ud{\a}{\b} \quad {\rm and} \quad
V\ud{\a}{\b} = -(\g^{\m})\du{\b}{\a}V_{\m}
\ee

 We also define the cousins of $\g^{\m}$ as follows:
\be \lb{cous}
(\left| \g^{\m}\right.)^{\a\b}=\ve^{\a\g}(\g^{\m})\du{\g}{\b}~,\quad
(\left. \g^{\m}\right|)_{\a\b}=\ve_{\b\g}(\g^{\m})\du{\a}{\g}~,\quad
(\bar{\g}^{\m})\ud{\a}{\b}=\ve^{\a\g}\ve_{\b\d}(\g^{\m})\du{\g}{\d}
\ee
The $\g$- and $\bar{\g}$-matrices are all traceless but not symmetric, whereas the
$\left. \g \right|$- and $\left| \g\right.$-matrices are all symmetric but not traceless.
Here are some useful identities:
\be \lb{iden4}
(\q\j)=(\j\q)~,\quad \q_{\a}\q_{\b}=\frac{1}{2}\ve_{\a\b}\q^2, \quad 
\q^{\a}\q^{\b}=-\frac{1}{2}\ve^{\a\b}\q^2
\ee 

\be \lb{iden5}
\q\g^{\m}\j =-\j\g^{\m}\q~,\quad (\q\j)(\q\f) = -\frac{1}{2}\q^2 (\j\f)
\ee

\be \lb{iden6}
(\q\x)\l_{\a} =\frac{1}{2}(\x\g^{\m}\l)(\g_{\m}\q)_{\a}-\frac{1}{2}(\x\l)\q_{\a}
\ee

\be \lb{iden7}
(\g^{\m}\q)_{\a}(\g^{\n}\q)_{\b}=\frac{1}{2}(\g^{\m}\left.\g^{\n}\right|)_{\a\b}\q^2
\ee

\be \lb{iden8}
(\g^{\m}\q)_{\a}(\g^{\n}\q)_{\b}\pa_{\m}\pa_{\n}=\frac{1}{2}\ve_{\a\b}\q^2\bo
\ee
where $\bo =\pa^{\m}\pa_{\m}$.

\section*{Appendix B: superspace notation~\footnote{Our notaton is different from that
 of ref.~\cite{sspace}.}}

3D superspace is parametrized by $(x^{\m},\q_{\a})$ where $\q$ is a Grassmann spinor.
As the book-keeping notation, we use
\be \lb{bookk2}
\pa_{\m}=\fracmm{\pa}{\pa x^{\m}}~~,\qquad
\pa_{\a}=\fracmm{\bvec{\pa}}{\pa \q^{\a}}
\ee 
It follows
\be \lb{eleme}
\pa_{\m}x^{\n}=\d^{\n}_{\m}~~,\qquad \pa_{\a}\q^{\b}=\d^{\b}_{\a}~~,\qquad
(\pa\pa)\q^2 = 4
\ee
Grassmann integration is equivalent to Grassmann differentiation,
\be \lb {grassi}
\int d\q_{\a}= \pa^{\a} \qquad {\rm and}\qquad \int d^2\q\,\q^2 =4 
\ee
The 3D supersymmetry generators are conveniently represented in 3D superspace by
\be \lb{susyge}
Q_{\a} = \pa_{\a} -i(\g^{\m}\q)_{\a}\pa_{\m} 
\ee
so that
\be \lb{susyal}
\{ Q_{\a}, Q_{\b} \} = 2(\left. \g^{\m} \right|)_{\a\b}P_{\m}~~,\qquad
\[ Q_{\a},P_{\m} \] =0
\ee
where $P_{\m}=i\pa_{\m}$ are the 3D translation generators. The 3D (flat) superspace 
supercovariant derivatives are defined by the relations
\be \lb{dal1}
\{ D_{\a}, Q_{\b} \} = \[  D_{\a}, P_{\m} \] =0 \quad {\rm and}\quad  
\{ D_{\a}, D_{\b} \} = 2i(\left.\g^{\m}\right|)_{\a\b}\pa_{\m}
\ee
It is easy to verify that
\be \lb{scovd}
 D_{\a} = -i\pa_{\a} + (\g^{\m}\q)_{\a}\pa_{\m}  
\ee
obey all eqs.~(\ref{dal1}). Here are some useful identities:
\be \lb{idd2}
D_{\a}D_{\b} = i(\left.\g^{\m}\right|)_{\a\b}\pa_{\m} +\frac{1}{2}\ve_{\a\b}D^2
\ee
\be \lb{idd3}
D^{\b}D_{\a}D_{\b} =0
\ee
\be \lb{idd4}
D^2D_{\a} = - D_{\a}D^2 = -2i(\g^{\m}D)_{\a}\pa_{\m}
\ee
\be \lb{idd5}
 (D^2)^2= 4 \bo 
\ee

\end{document}